\documentclass[referee]{mn2e}
\usepackage{graphics}
\title[Carriers for the 21- and 30-$\mu$m nebular bands]{Candidate carriers and synthetic spectra of the 21- and 30-$\mu$m proto-planetary nebular bands}
\author[R. Papoular]{R. Papoular$^{1}$\thanks{E-mail:
papoular@wanadoo.fr}\\
$^{1}$Service d'Astrophysique and Service de Chimie Moleculaire,\\
CEA Saclay, 91191 Gif-s-Yvette, France}
\begin{document}

\date{Accepted . Received ; in original form }

\pagerange{\pageref{firstpage}--\pageref{lastpage}} \pubyear{2002}

   \maketitle
\label{firstpage}

\begin{abstract}

Computational chemistry is used here to determine the vibrational line spectrum of several candidate molecules. It is shown that the thiourea functional group, associated with various carbonaceous structures (mainly compact and linear aromatic clusters), is able to mimic the 21-$\mu$m band emitted by a number of proto-planetary nebulae. The combination of nitrogen and sulphur in thiourea is the essential source of emission in this model: the band disappears if these species are replaced by carbon.

The astronomical 21-$\mu$m feature extends redward to merge with another prominent band peaking between 25 and 30 $\mu$m, also known as the 30-$\mu$m band. It is found that the latter can be modelled by the combined spectra of aliphatic chains, made of CH$_{2}$ groups, oxygen bridges and OH groups, which provide the 30-$\mu$m emission. The absence of oxygen all but extinguishes the 30-$\mu$m emission. The emission between the 21- and 30-$\mu$m bands is provided mainly by thiourea attached to linear aromatic clusters.

The chemical software reveals that the essential role of the heteroatoms N, S and O stems from their large electronic charge. It also allows to determine the type of atomic vibration responsible for the different lines of each structure, which helps selecting the most relevant structures.

Obviously, no single structure can exhibit the required spectrum, for each only contributes discrete lines which cannot be broadened enough by usual mechanisms. A total of 22 structures have been selected here, but their list is far from being exhaustive; they are only intended as examples of 3 generic classes. However the concatenation, interpolation and smoothing of the computed line spectra deliver continuous spectra of the overall emission of the selected candidates. \bf When  background dust emission is added, model spectra are obtained, which are able to satisfactorily reproduce recent observations of proto-planetary nebulae. \rm

The relative numbers of atomic species used in this model are typically H:C:O:N:S=53:36:8:2:1.

\end{abstract}




\section{Introduction}

Since their discovery of the 21-$\mu$ line emitted by proto-planetary nebulae (PPNe) in the IRAS data bank, Kwok and coll. \cite{kwo89} have added many observational contributions to the knowledge of PPN spectra: see the bibliography in Zhang et al. \cite{zha}. We shall dwell here on this paper, which reports on their Spitzer/IRS study of 21 and 30 $\mu$m by several galactic PPNe. Successive observational improvements have led to pinpoint the former's peak wavelength at 20.1 $\mu$m
 (Hrivnak et al. \cite{hri}).

 PPNe have a relatively short lifetime between the stellar states of AGB (Asymptotic Giant Branch) and PN (Planetary Nebula), and the number of available sources is correspondingly small. Nevertheless, the position of the 21-$\mu$m band is very accurately determined, as is its shape, slightly skewed, with a red wing more extended than the blue one. On the other hand, the amplitude of the 30-$\mu$m band relative to the former is widely variable and its  shape does not seem to be stable; it extends from about 23 to about 40 $\mu$m, and peaks between 26 and 35 $\mu$m.

The present work dwells on possible carriers of these two bands. Over the years, several candidates were put forth. For the 30-$\mu$m band, Goebel \cite{goe85} initially proposed MgS, and was followed by Szczerba et al. \cite{szc} and Hony et al. \cite{hon}. Omont et al. \cite{omo} concluded that this choice is disputable. Papoular \cite{pap00} later suggested a strong contribution from hydroxyl groups attached to various carbonaceous structures.

For the 21-$\mu$m band, Goebel \cite{goe93}, for instance, suggested solid SiS$_{2}$, and solid particles of SiC in various shapes and sizes were advocated by Speck and Hofmeister \cite{spe}). Posch et al. \cite{pos} published experimental and theoretical spectra of SiO$_{2}$-coated SiC and FeO, which exhibit very convincing strong features near 20 $\mu$m. Von Helden et al. \cite{hel} advanced small clusters of titanium carbide with a special composition, Ti$_{14}$C$_{13}$, one of several others experimentaly studied by v. Heijsnbergen, v. Helden et al. \cite{hei}. Each of these candidates has, of course, its pros and cons, which were discussed in the same literature; see in particular the extensive and unbiased discussion by  Posch et al. \cite{pos}. However, TiC attracted special attention in view of the excellent agreement, in position and width, of the published laboratory feature with the PPN feature. For the same reason, several authors looked further into this candidate. Henning and Mutschke \cite{hen} showed that the feature is not present in the reflection spectrum of bulk TiC, and deduced from their measurements that it could not be exhibited by small TiC particles of various shapes. Li \cite{li} ruled out the TiC model based on the excessive amounts of TiC required by Kramers-Kroenig relations between total dust mass and integrated extinction cross-section. The same conclusion was arrived at by Chigai et al. \cite{chi}. Zhang et al. \cite{zhak} also critically analyzed several of the published proposals.

Thiourea was already proposed as a carrier of the 21-$\mu$m feature by Sourisseau et al. \cite{sou}.  It is a common molecule on earth and in the industry, and was described in detail by Stewart \cite{ste}, Kutzelnigg and Meck \cite{kut}, Lin-Vien \cite{lin} and Alia et al. \cite{ali}, and in Ullman's Encyclopedia of Industrial Chemistry \cite{ull} among others (see further relevant bibliography in Sourisseau et al. \cite{sou}). Its chemical formula is SC(NH$_{2})_{2}$. As an aside, it differs from urea only by the presence of sulphur, S, instead of oxygen, O. Its characteristic subgroup, CS, was observed along several sightlines in the Galaxy (e.g. Turner \cite{tur87b}). Also note that the cosmic abundance of sulphur is slightly higher than that of Silicon, which is known to associate readily with carbon, both being chemically active. Sulphur associated with silicon was also detected in the form of SiS molecules (e.g. Turner \cite{tur87a}).

\begin{figure}
\resizebox{\hsize}{!}{\includegraphics{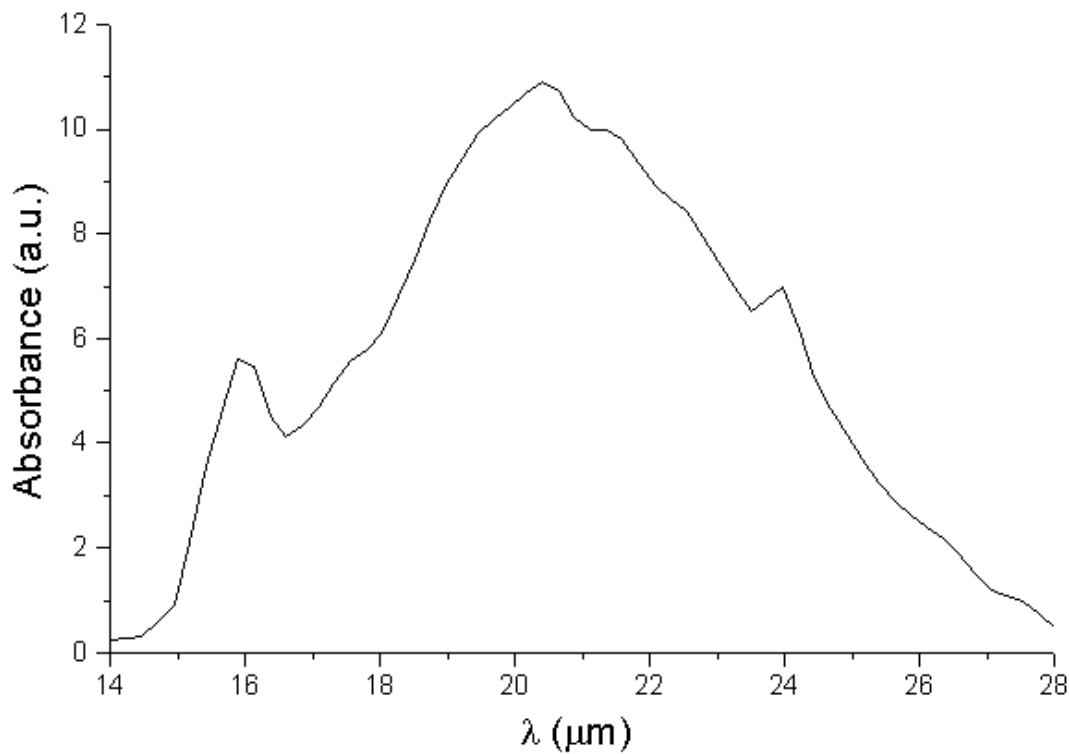}}
\caption[]{The absorbance of solid thiourea prepared in a Nujol mull. Note the main peak at 20.4 $\mu$m, a secondary peak at 15.8 $\mu$m, and an elbow at 24 $\mu$m. FWHM$\sim$7 $\mu$m.}
\end{figure}

According to Stewart \cite{ste}, thiourea exhibits an absorption band which is reproduced in Fig.1; it peaks at 20.4 $\mu$m, but is distinctly wider than the PPNe feature, and much wider than molecular features, for that matter. The thiourea molecule, being made of 8 atoms, has 18 vibrational modes; theoretical analysis indicates that only 3 of them are notably IR-active between 10 and 40 $\mu$m, and the strongest falls indeed at 486 cm$^{-1}$ (20.6 $\mu$m). But this alone cannot be responsible for the band width in Fig.1. Obviously, this is mainly due to its preparation, as solid state micrograins, embedded in a Nujol mull (which is expected to add a small, uniform absorbance). It is well known that the spectrum of small grains is different from that of the same material in bulk, and that particle size and shape have an enormous influence on the position, shape and width of spectral lines (see Smith \cite{smi}) . A popular example, of course, is SiC, which has a very discrete mid-IR spectrum with a single Lorentz oscillator at $\nu_{t}$=795 cm$^{-1}$. This single oscillator is responsible for a great variety of measured spectra, depending on particle specifics (see Bohren and Huffman \cite{boh}, Chap. 12). Therefore, if thiourea is to be considered, it could not be in the solid state.

  Now, the single, unbroadened line of the gaseous molecule near 20 $\mu$m cannot mimic the 21-$\mu$m band, whose width is 2.4 $\mu$m. However, broadening of the line may also occur due to the staightforward formation of H-bonds and complexes with impurity ions (coordination) or to the presence of thiourea derivatives (see Stewart \cite{ste}, Masunov and Dannenberg \cite{mas} Bryantsev and Hay \cite{bry}, Brennan \cite{bre}). These are of the form SC(NR1R2)(NR3R4), where the R's represent radicals attached to the SCN$_{2}$ root.

The present work, precisely, explores the possibility of taking advantage of this propensity of thiourea in order to control the position and shape of its 21-$\mu$m feature so as to better fit the observations. Indeed, attachment to a chemically different structure slightly shifts the thiourea lines and may also enhance the IR activity of some lines nearby. Combining the spectra of many associations of this type will hopefully fill the window occupied by the astronomical band. In chosing the structures to be associated with the thiourea group, consideration of cosmic abundances shows that simple hydrocarbons may be considered as useful radicals for our purpose.

The astronomical 21-$\mu$m feature extends redward to merge with another prominent band peaking between 25 and 30 $\mu$m, also known as the 30-$\mu$m band. Following the same line of thought, it is found that this ubiquitous band can be modelled by the combined spectra of a large number  of aliphatic chains, made of CH$_{2}$ groups, oxygen bridges and OH groups.

The experimental literature on thiourea and its  derivatives in the gas phase is scant, despite a recent surge of interest spured by possible chemical applications (see Lesarri \cite{les}). It is therefore necessary to resort to theory and computational chemistry (see Alia et al.\cite{ali}, Masunov et al. \cite{mas}, Bryantsev et al. \cite{bry}, etc.). Fortunately, the latter has progressed considerably in the last decade, especially due to the greatly increased speed of computers. The procedure followed in the present work is the same as that which was described in detail in Papoular (\cite{pap01}), except that the adopted chemical software was updated to version 7 of Hyperchem (Hypercube, Inc.).

This software delivers, for each mode, the IR intensity and graphic illustration of the movement of each atom in the structure, together with the frequency of vibration. This is of great help in selecting chemical elements and molecular structures of interest, and later estimating their relative abundances in the model.

The semi-empirical, PM3/RHF computation method was preferred for the aliphatic structures because it was specifically optimized for hydrocarbon structures and gives sufficiently accurate IR freqencies (better than about 5 $\%$) within reasonable computation times. For the thiourea family, the semi-empirical AM1 method was chosen for its more accurate treatment of N and S atoms.

Sections 2 and 3 respectively deal with the 2 generic classes of molecules defined above. In each case, several examples of structures are illustrated, together with the corresponding IR spectra, and, whenever possible, the type of molecular vibration is described. 

In Section 4, we synthesize all the above line spectra and  exhibit typical emission spectra, differing in the temperatures and abundances of the emitters, to be compared with typical observations. The required relative abundances of the various atomic species are indicated in Sec. 5.

\section{The thiourea family}

The theoretical fundamental vibrational spectrum of thiourea is obtained by first optimizing the structure of the molecule in its neutral ground state, then performing a normal mode analysis. This delivers the vibrational frequencies and geometries in the limit of weak excitation or low temperatures.

\begin{figure}
\resizebox{\hsize}{!}{\includegraphics{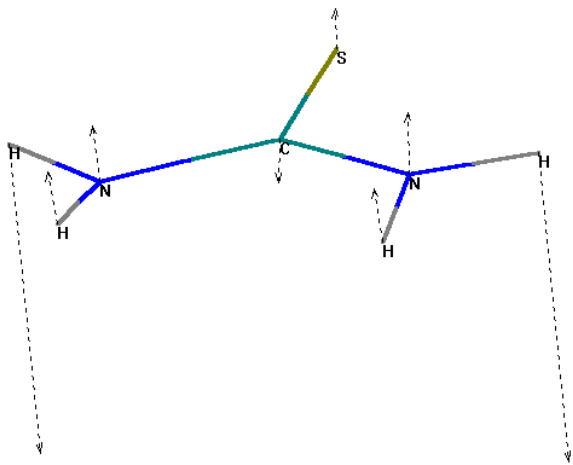}}
\caption[]{The thiourea molecule in its most stable state: slightly skewed out of plane, -662 kcal/mol, symmetry C$_{2}$ (anti). The dashed arrows represent the excursions of the atoms vibrating in the mode of interest, at 20.3 $\mu$m.}
\end{figure}

The optimized thiourea molecule (delivered by our computations) is drawn in Fig.2, in its most stable state, where the molecule is shown to be (only) nearly plane. 
According to Lesarri et al. \cite{les}, the experimental values for the CS bond length, the CN bond length and the SCN angle are, respectively, 1.645 \AA{\ }, 1.368 \AA{\ }, and 123 $\deg$, all for the ground state. The corresponding values obtained in the present work are 1.63, 1.38 and 120.5 respectively.

The vibrational mode of interest here is illustrated by dashed arrows in Fig. 2. The line frequency computed by the software is 492 cm$^{-1}$; the corresponding value obtained by Stewart \cite{ste} and Yamagushi et al. \cite{yam} are 486 and 498, respectively. Appendix A shows the extent of aggreement over the IR spectrum, between Stewart's, Yamagushi's and the present results. 

\begin{figure}
\resizebox{\hsize}{!}{\includegraphics{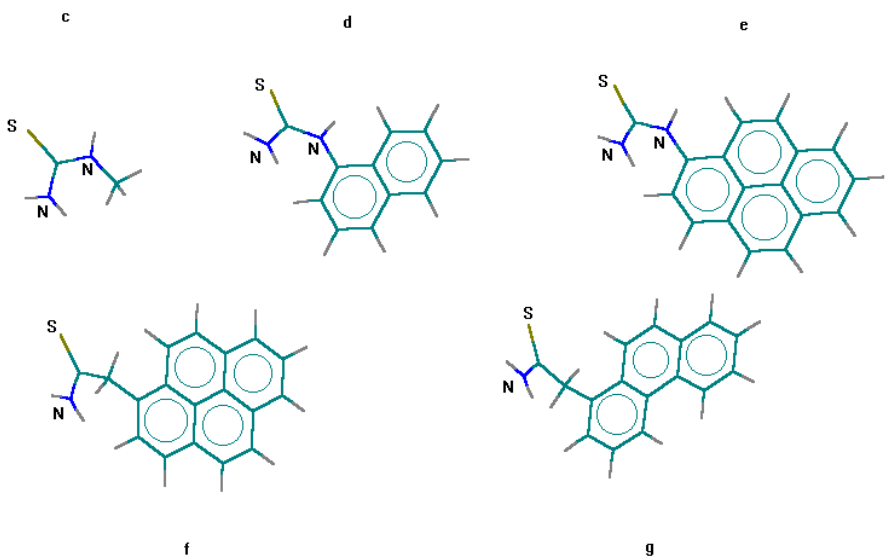}}
\caption[]{A set of thiourea derivatives that were selected among many others for present purposes. Conforming to chemical conventions, only the heteroatoms N (blue) and S (yellow) are explicitely labeled, while C's are coloured in green and H's in gray. In this set, thiourea is appended to mainly compact aromatic structures. In the (c,d,e) structures, both the N atoms of thiourea are conserved, and one of them loses one of its H's. Structures f and g have only one N, but that does not preclude their contributing to the 20 $\mu$m spectrum.}
\end{figure}

Considering thiourea as a functional group, it is reasonable to conjecture that its characteristic lines should survive attachment to different chemical structures, designated by R's in its general formula, SC(NR1R2)(NR3R4). Such associations are expected to slightly shift the thiourea lines to and fro, which could help filling the spectral window covered by the observed 21 $\mu$m band. In space, of course, the most likely candidates for R's are hydrocarbons. As examples of possible associates to thiourea, Fig. 3  collects mainly associations with compact aromatic clusters, while Fig. 5 assembles cases of linear clusters.

\begin{figure}
\resizebox{\hsize}{!}{\includegraphics{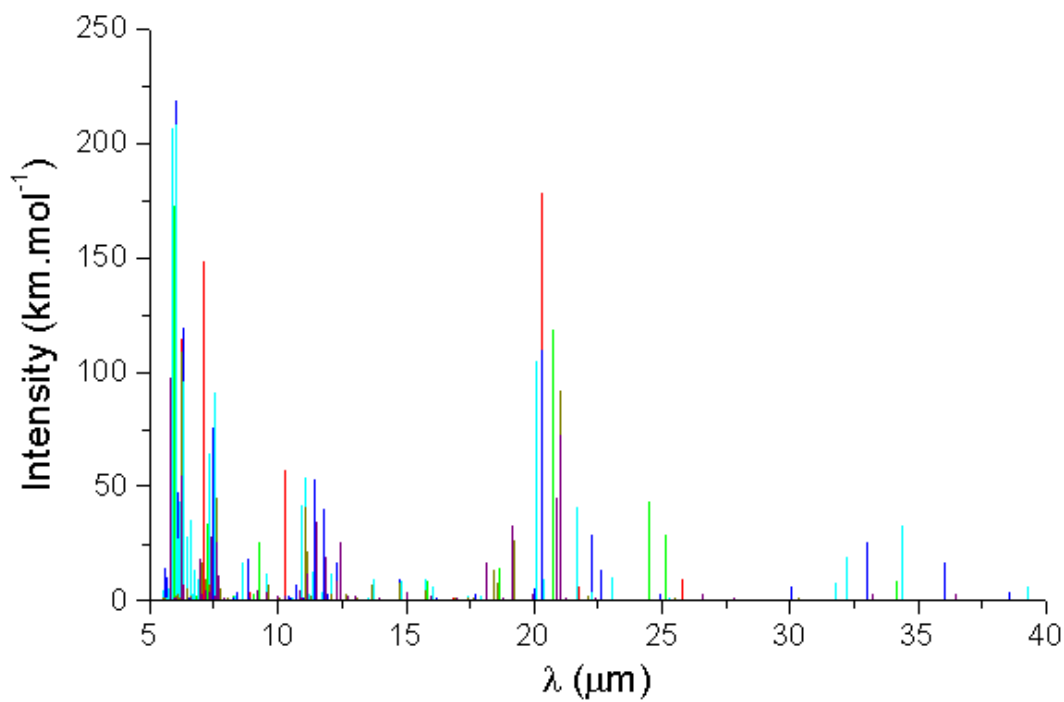}}
\caption[]{Intensity spectra computed for bare thiourea in red, and its selected derivatives in Fig. 3 (involving mainly compact aromatic structures): c) green, d) blue, e) cyan, f) dark yellow, g) purple. All structures contribute to the 21-$\mu$m band, between 18 and 23 $\mu$m, peaking at 20.3 $\mu$m; some also contribute (moderately) between 30 and 40 $\mu$m.}
\end{figure}

\begin{figure}
\resizebox{\hsize}{!}{\includegraphics{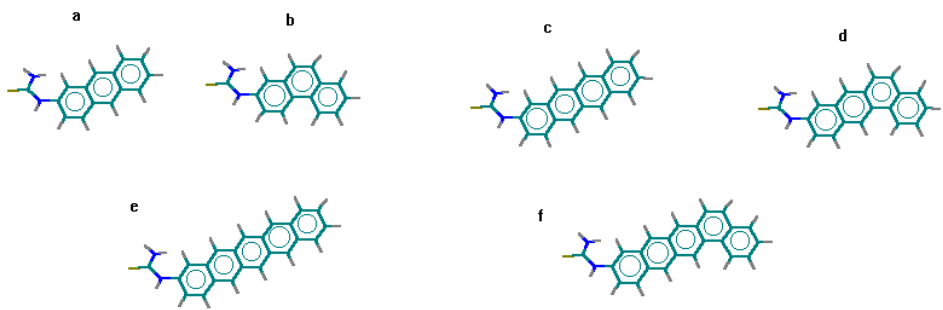}}
\caption[]{Selected thiourea derivatives involving linear aromatic structures.}
\end{figure}

\begin{figure}
\resizebox{\hsize}{!}{\includegraphics{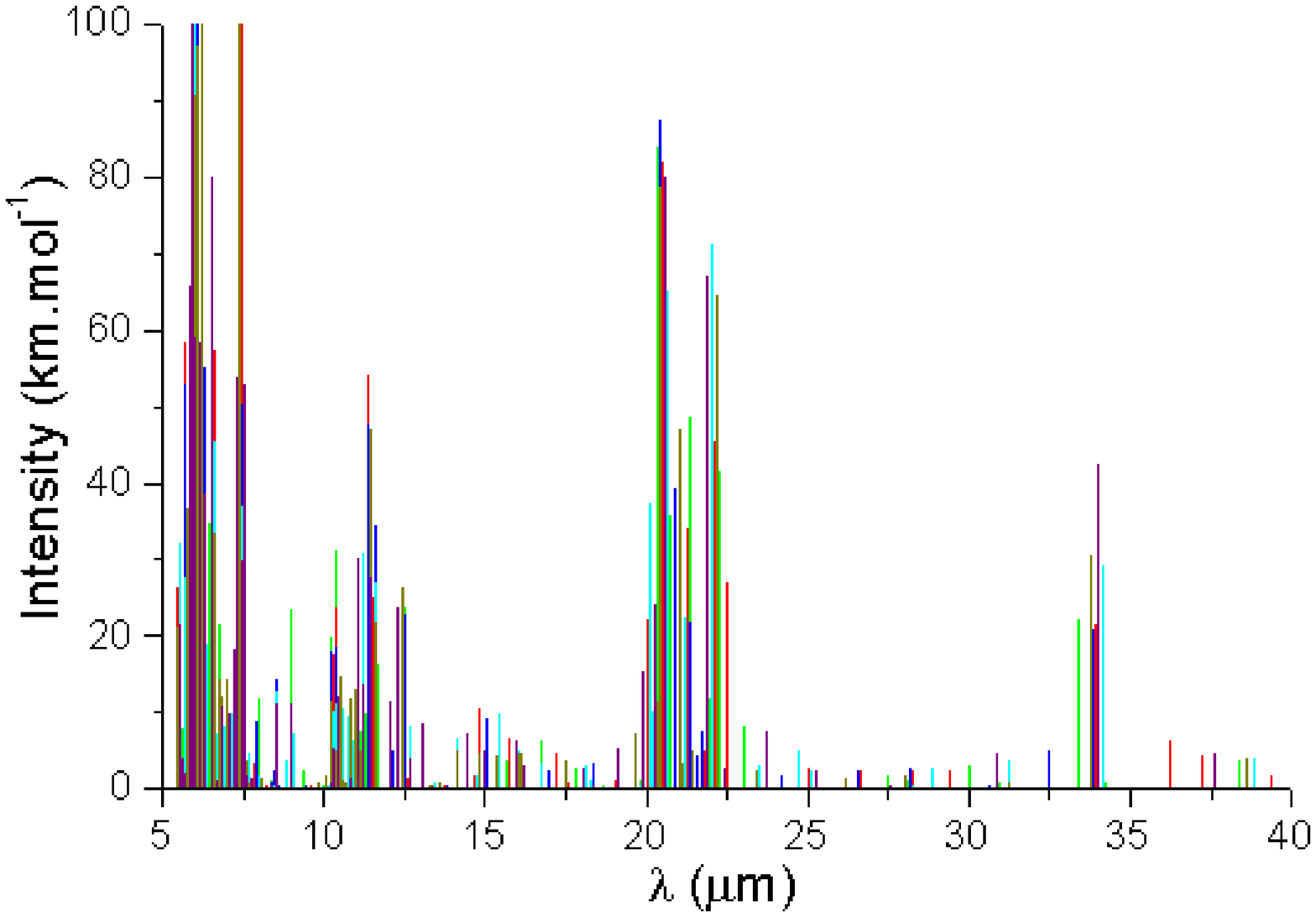}}
\caption[]{Intensity spectra computed for the set of thiourea derivatives in Fig. 5. Respectively, a)red, b) green, c) blue, d) cyan, e) dark yellow, f) purple. Here, the peak intensities are half as strong as in Fig. 4, but the main bunch of lines of interest, near 20 $\mu$m, extends farther, to 23 $\mu$m.}
\end{figure}

 In each derivative, the thiourea root, with both N's or only one N left, is attached to a more or less complicated hydrocarbon structure. The spectral lines of the two groups of structures are represented in Fig. 4 and 6, respectively, by vertical lines of different colors, and lengths proportional to their computed integrated band intensities, $I$, in units of km.mol$^{-1}$, given by

\begin{equation}
I=\frac{\alpha\Delta\nu}{10^{2}\,C}
\end{equation}

where $\alpha$ is the corresponding absorbance in cm$^{-1}$, $\Delta\nu$ is the band width in 
cm${-1}$, and $C$ is the molecular density in mol.l$^{-1}$ of the sample used for measuring $\alpha$.

Graphic analysis of the modes shows that the thiourea family of Fig. 4 contributes lines essentially from 18 to 23 $\mu$m. By usual standards of IR activity, the first group is particularly strong. The lines close to 20 $\mu$m are due to  symmetric out-of-plane (oop) excursions of the upper NH's (Fig. 1) and little oop excursions of S. The lines beyond are contributed by molecules carrying 3 NH bonds in asymmetric oop excursions of the NH's and ip or oop excursions of S. 

The spectra of thiourea appended to linear aromatics are represented in Fig. 6 in different colours. As compared to those of Fig. 4, they are roughly half as strong, but they extend more uniformly from 20 to 22.5 $\mu$m, which will help building the extended red wing characteristic of the astronomical 21-$\mu$m band, as will be seen shortly. Note that the lines of the same, but bare, aromatics, flock around 22 $\mu$m. They are due to orderly bulk oop vibrations, akin to string vibrations and their frequency depends on the relative phases of the carbon atoms excursions. As a rule vibrations of the straight leg of the structure contribute to the spectrum near 22 $\mu$m, while the other leg contributes near 26 $\mu$m. The association with the thiourea chemical group activates and shifts some of the thiourea lines. Hence the increased line crowding near 20 and 22 $\mu$m.

Apart from helping shift the 20-$\mu$m line, the main contributions of the attached hydrocarbon sub-structures are in the ranges of the near- and mid-IR UIBs (Unidentified IR Bands) observed in the sky, to which, of course, the thiourea functional group also contributes (essentially with NH stretching and ip (in-plane) and oop motions).

Even when concatenated, the spectra of Fig. 4 and 6 are obviously still too sparse to make a continuous band. The structures illustrated in Fig. 3 and 5 are only intended as examples of generic molecules from which to generate other carriers of the 21-$\mu$m line. This may be done by varying the number of attached phenyl rings, or by changing the position of the thiourea group around the periphery. Leaving this for further work, we are content here with a smoothing operation consisting of Fast Fourier Transform smoothing over 30 points (this procedure is more effective than the usual adjacent averaging device). 

\begin{figure}
\resizebox{\hsize}{!}{\includegraphics{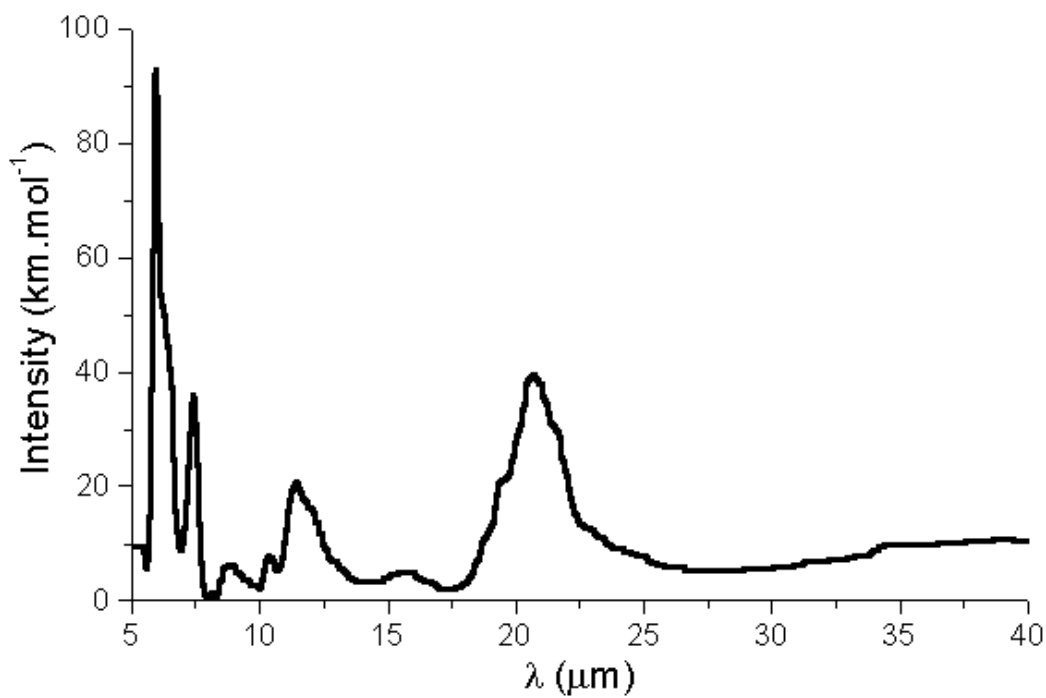}}
\caption[]{The smoothed concatenation of the line spectra of Fig. 4 and 6, due to thiourea and its selected derivatives in Fig. 2, 3 and 5. The central peak is at 20.7 $\mu$m; FWHM$\sim2.3\mu$m,  to be compared, respectively, with  20.1 and 2.4 $\mu$m, as determined by Hrivnak et al. \cite{hri}. Note its trailing red wing, mainly due to thiourea derivatives involving aromatics in rows; also  note the bands at 11.5 and 15.5 $\mu$m.}
\end{figure}

The central peak is at 20.7 $\mu$m, to be compared with 20.4 $\mu$m, for the experimental absorbance in Fig. 1. The much broader width of the latter may be explained along the lines  suggested in the introduction. The points at half-height in Fig. 7 are at 19.6 and 21.9 $\mu$m, to be compared with 19.4 and 21.8, on average, as determined by Hrivnak et al. \cite{hri} for the 21-$\mu$m line. However, the peak wavelength is shifted redward from 20.1 by 0.6 $\mu$m, closer to 21 $\mu$m. This discrepancy remains to be resolved.

\section{The aliphatic family}
The 30-$\mu$m feature roughly extends from 23 to 40 $\mu$m. While the thiourea family of molecules makes some contribution to this range, this is only a minor one. It is therefore necessary to look for another plausible carrier.

\begin{figure}
\resizebox{\hsize}{!}{\includegraphics{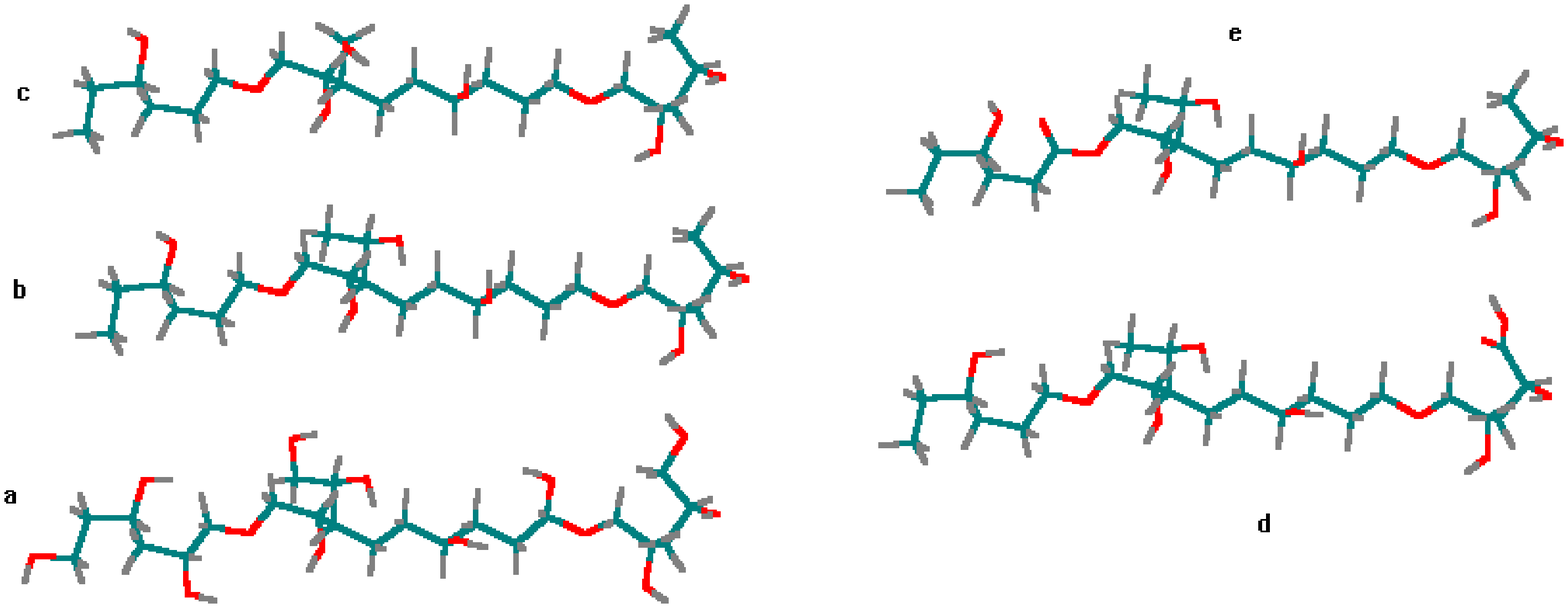}}
\caption[]{A family of aliphatic chains that were selected as examples of candidate carriers of the 30-$\mu$m PPN band. Oxygen atoms are couloured in red. The relative number of attached OH groups is much higher than expected in space, considering the relative cosmic abundance of O; but this highlights the role of hydroxyls, which, in space, may be distributed among many more different chains than can be handled here.}
\end{figure}

\begin{figure}
\resizebox{\hsize}{!}{\includegraphics{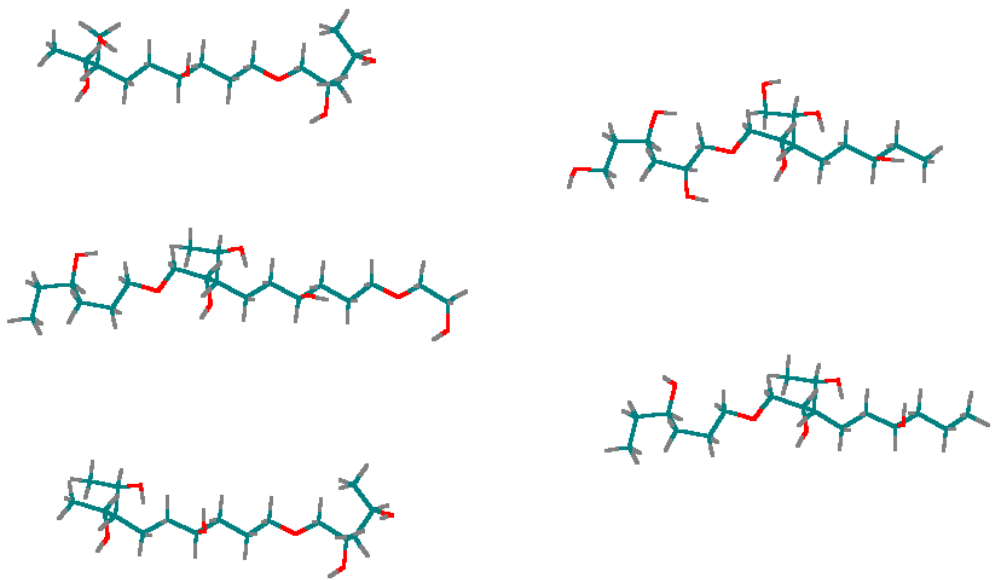}}
\caption[]{Another, analogous set of aliphatic chains that were selected as examples of candidate carriers of the 30-$\mu$m PPN band.}
\end{figure}

\begin{figure}
\resizebox{\hsize}{!}{\includegraphics{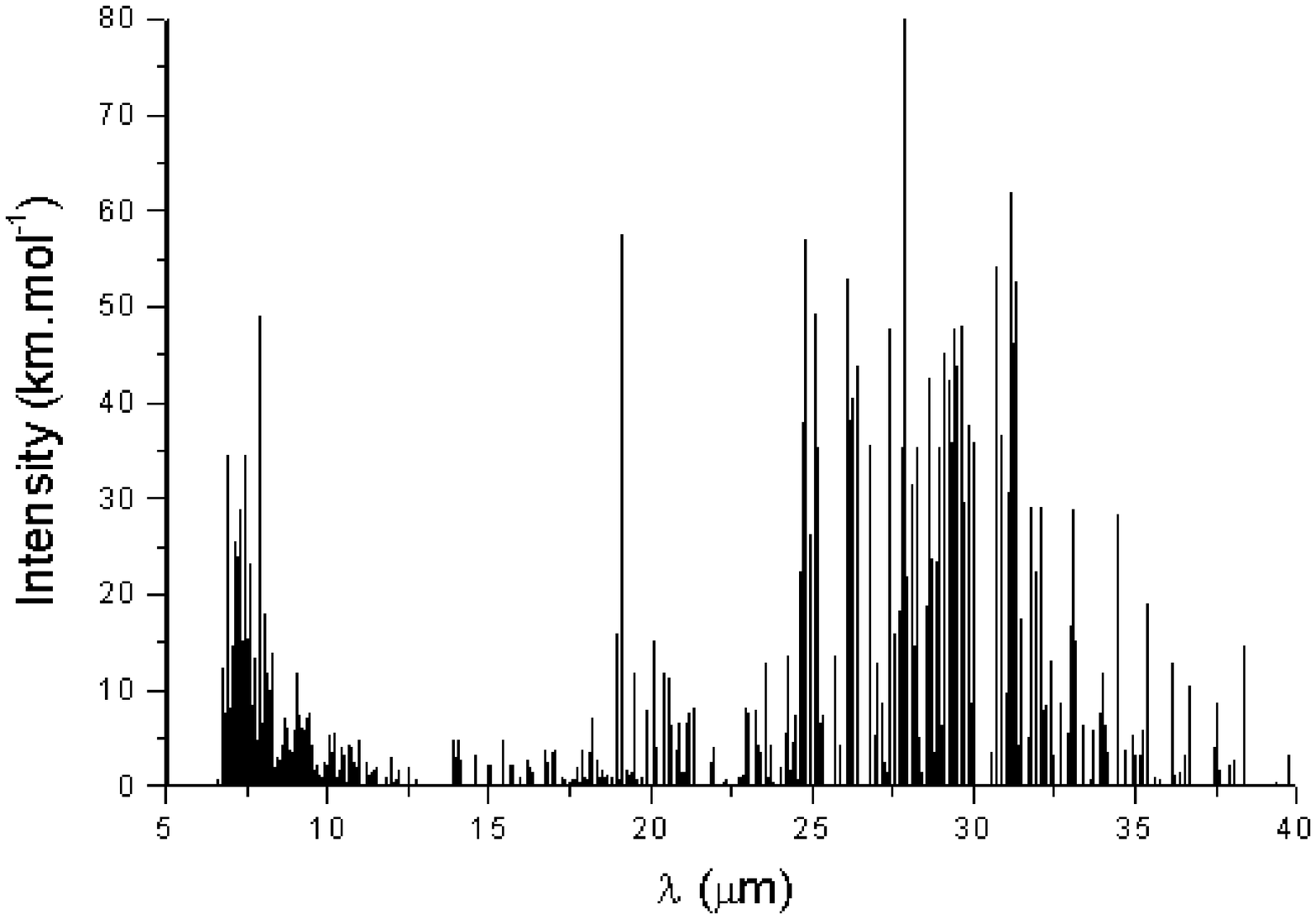}}
\caption[]{The concatenated intensity spectrum of the aliphatic family illustrated in Fig. 8 and 9. Note here the much weaker contribution to the near and mid-IR, than in the case of the thiourea derivatives, which include compact aromatics.}
\end{figure}

 It was shown in Papoular \cite{pap} that the OH radical (hydroxyl) considerably increases the IR activity of the structures to which it is attached, especially in the range 30 to 40 $\mu$m. This, again, is because of the strong electric charge of the oxygen atom. The oop, wagging motion of the OH bond, which has a strong IR intensity, is the main contributor to this range. Chain geometry is especially favourable to OH attachment. Note that, in the kerogen model of UIB carriers, aliphatic chains are expected to link other, compact structures, and contribute essentially to the 8.6-$\mu$m band (Papoular \cite{pap}). The spectral effect of OH attachment is best illustrated in the case of linear aliphatic chains, essentially made of CH$_{2}$ groups, connected by single CC bonds. These have intrinsically only a very weak IR activity. However, when OH groups are attached, as in Fig. 8 and 9, strong lines arise, mainly in the range 20 to 40 $\mu$m, as shown in Fig 10.

 Apart from the lines of interest between 22 and 27 $\mu$m, the CH stretching , ip and oop vibrations, as might be expected, contribute several intense lines within the UIBs. Again, this series of structures can be continued along the same line to obtain a denser concatenated spectrum.

\section{Synthesis of spectra}

\bf We are now in a position to try and simulate typical PPN spectra by combining the 2 thiourea sub-groups and the aliphatic group of structures defined above. In particular, we wish to obtain a spectrum with nearly equal emission at 20 and 30 $\mu$m (e.g. IRAS source 23304+6147), and a spectrum where the latter is much stronger than the former (e.g. IRAS 22574+6609). 

Note that Zhang et al. \cite{zha} , in fitting their latest Spitzer/IRS PPN spectra, also needed 3 Lorentz features to account for the same FIR spectral window. They also added, as a background, 2 dust continua at different temperatures. Following suit, let us seek fitting combinations of molecular and background dust temperatures and intensities. The temperatures of our 3 molecular families will be assumed equal ($T_{m}$), but not necessarily equal to those of the dust ($T_{d1}$ and $T_{d2}$).

Consider, first, IRAS 22574+6609. Based on the line densities and intensities of our 3 families of structures (Fig. 4, 6 and 7), we shall assume that thiourea derivatives (with compact and linear aromatics) and the aliphatic chains (Fig. 2, 3, 5, 8 and 9) are in the ratios 1:1:1. Thus, in Fig.11, we concatenate the spectra of Fig. 4 (red) and 6 (green) together with the spectrum of Fig. 10 (blue).

\begin{figure}
\resizebox{\hsize}{!}{\includegraphics{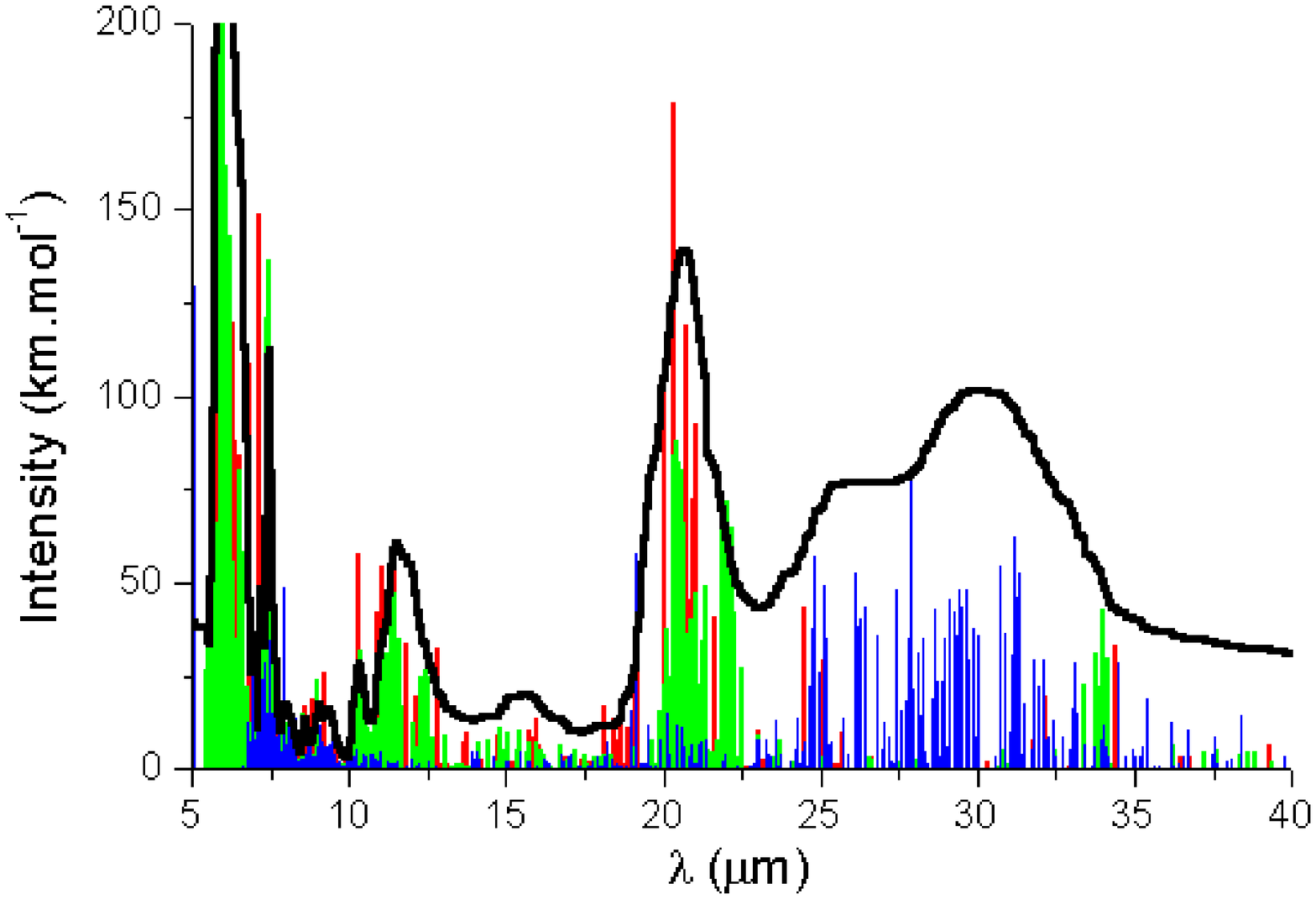}}
\caption[]{Columns; red: the concatenation of the spectral lines (height in km.mol$^{-1}$) of thiourea and thiourea derivatives with compact aromatics; green: thiourea derivatives with linear aromatics; blue: aliphatic chains with attached OH chemical groups. Black continuous line: the result of Fast Fourier Transform smoothing (an improved version of Adjacent Averaging), and multiplication by 5 for clarity. The 21-$\mu$m feature peaks at 20.7$\mu$m; FWHM=2.3 $\mu$m. Note the shoulder at 26 $\mu$, reminiscent of the 26-$\mu$m spectral component conjectured by Kwok andcollaborators.}
\end{figure}

Because we have not considered enough individual structures in each family, the concatenated spectrum of all the selected structures is still not nearly continuous. We therefore need some sort of interpolation and/or smoothing. Accordingly, the concatenated line spectrum was Fast Fourier Transform smoothed  over 50 points, resulting in the continuous black line in Fig. 11. The 20-$\mu$m feature peaks at 20.7$\mu$m; its apparent FWHM is 2.3 $\mu$m.

Since the integrated band intensity, $I$, is proportional to the absorbance, $\alpha$, of the material (eq. 1), its radiative emission, $F_{\lambda}$, will be proportional to the product of $I$ and the black body emission at the temperature, $T_{m}$, of the molecules. This will be multiplied by $\lambda$, for comparison with the $\lambda\,F_{\lambda}$ spectra of Zhang et al. \cite{zha}. For the same reason, we adopt dust emissivities scaling like $\lambda^{-2}$.

The fitting procedure delivered the spectrum of Fig. 12 (continuous curve): $T_{m}$=150 K, $T_{d1}$= 170 K, $T_{d2}$=67 K; to be compared with the spectrum of IRAS 22574+6609 (overlaid, dots). No attempt was made in the model to account for the UIBs, which are observed to be prominent between 11 and 13 $\mu$m. Although the thiourea family is seen to make a contribution to this range (Fig. 11), its intensity and temperature are too low for it to emerge in Fig. 12.

\begin{figure}
\resizebox{\hsize}{!}{\includegraphics{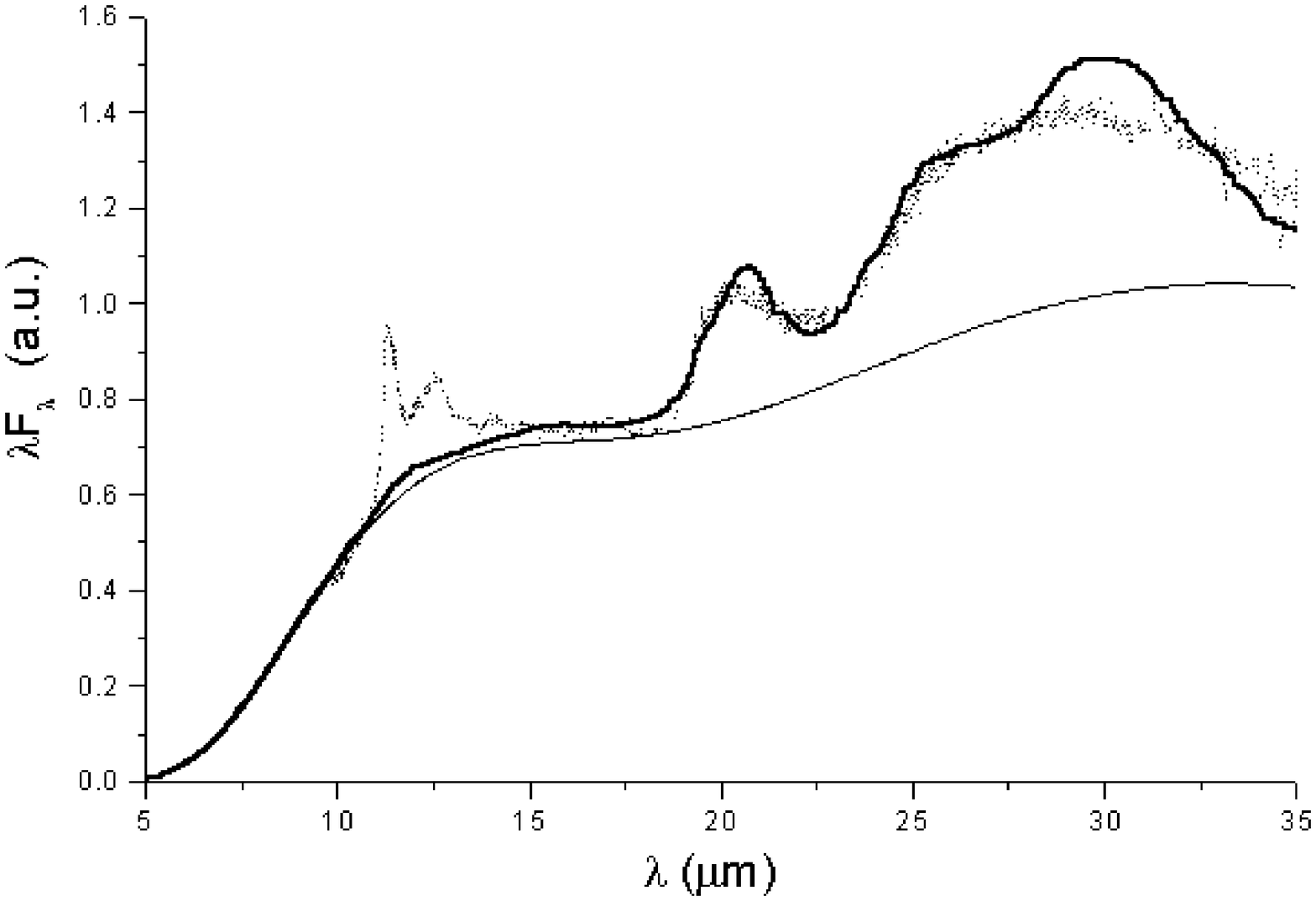}}
\caption[]{The measured spectrum of IRAS 22574+6609 (dots); a model for it (thick continuous line): $T_{m}$ (molecules)=150 K, $T_{d1}$ (dust 1)= 170 K, $T_{d2}$ (dust 2)=67 K. The sum of the two dust contributions is also drawn (thin line). The mid-IR component is not prominent here, because the model UIB carriers are not included (see Papoular \cite{pap}). }
\end{figure}

Now to IRAS 23304+6147. Here, we found it necessary to reduce the abundances of the two thiourea families relative to that of the aliphatic chains, so the ratios became 0.67:0.67:1. The second dust temperature was also changed into $T_{d2}$=78 K, and the abundance of molecules relative to dust was increased by a factor $\sim3.3$. The measured and model spectra are drawn in Fig. 13, and the same comment as for Fig. 12 can be made as to the intensities in the UIB range. Note that, in this second case, the 16-$\mu$m band appears in both the original and the model.

\begin{figure}
\resizebox{\hsize}{!}{\includegraphics{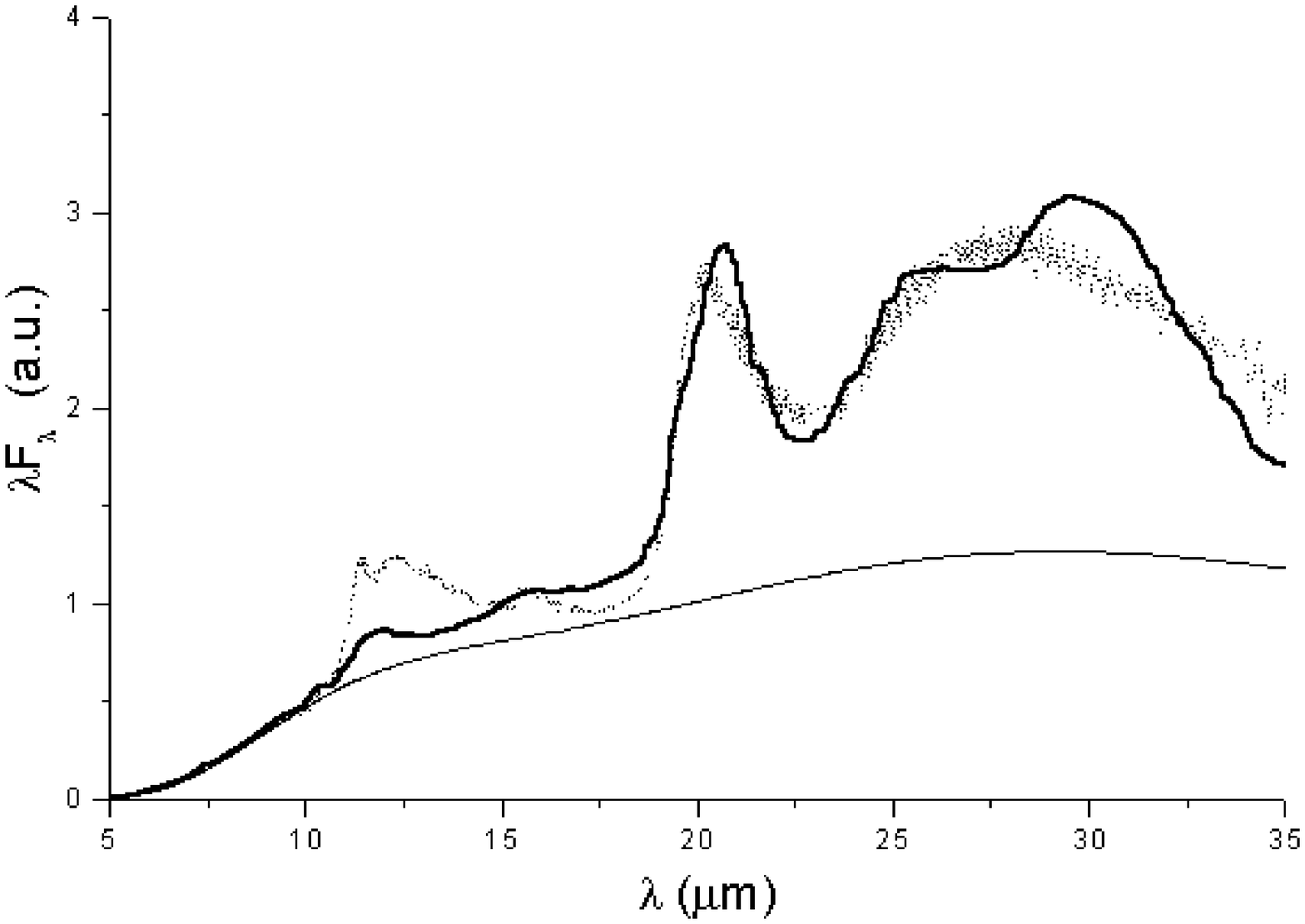}}
\caption[]{The measured spectrum of IRAS 23304+6147 (dots); a model for it (thick continuous line): $T_{m}$ (molecules)=150 K, $T_{d1}$ (dust 1)= 170 K, $T_{d2}$ (dust 2)=78 K. The sum of the two dust contributions is also drawn (thin line). The mid-IR component emerges but is still not prominent, because the model UIB carriers are not included (see Papoular \cite{pap}). Here, however, the 16-$\mu$m band appears in both the original and the model. }
\end{figure}

There is clearly a possible trade-off between temperatures and relative abundances of the 3 families of molecules, as well as the dust temperatures and abundances, so the choices made above are not unique. The resolution of this ambiguity requires a knowledge of the excitation process of the molecules, and of their interactions with local dust.

An obvious defect of these model spectra is the intensity dip near 27 $\mu$m (Fig. 11), which distorts the band top and shifts the peak to $\sim30\mu$m in Fig. 12 and 13. \rm  Correction of such defects requires a more elaborate apportion of the various selected structures, or the addition of new ones. An example for the latter could be structure (e) in Fig. 6, where 2 OH groups happen to be nearly parallel and vibrate in phase, giving the strongest line in Fig. 10, at 27.9 $\mu$m.

\bf Another defect is the shift of the model peaks redward from 20.1 to 20.7 $\mu$m. This makes a relative mismatch of 3$\%$, well within the wavelength error margin of our chemical software. However, as noted earlier, the respective wavelength intervals occupied by the model FIR bands comply with observations, and the wavelengths at half-maximum of the bare 21-$\mu$m band (Fig. 7) nearly coincide with those of the normalized band determined by Hrivnak et al. \cite{hri}. Rather than the wavelengths, one is therefore led to question the adopted relative structure abundances, or intensity errors (due to intrinsic inaccuracy of the software, and larger than wavenumber errors) which are more likely to affect the band profile and, hence, the peak position. Thus, the lines of thiourea and thiourea derivative (e) in Fig. 3, fall at 20.3 and 20.1 $\mu$m, respectively; if their relative abundances or their intrinsic IR intensities were underestimated in the models, any increase in their values would shift the band peak in the right direction. It should also be worth looking into the details of extracting the 21-$\mu$m band from the full spectra, and inquiring if these may affect the peak position.\rm 

\section{Elemental composition of the carriers}

Table 1 lists the various structures with their elemental compositions for each family. \bf The last two lines pertain to case 1 (Fig. 12) and give the global composition of the model carriers. The global ratio C/H is characteristic of predominently aliphatic structures. The roles of N and S, on the one hand, and O (mainly in OH) on the other hand, in building the model spectra, are made clear by Fig. 11: in this model, they are responsible for the 21- and 30-$\mu$m bands, respectively. These heavy elements are among the most abundant in the circumstellar medium; they have been detected in a large number of molecules (e.g. CO, NO, CS HNCS, OCS, etc.). We recall that the cosmic abundances of O, N and S relative to C are of order 1, 0.1 and 0.1, respectively. By comparison, the small relative amounts required by our model are expected to be easily available in the PPN environment.

On the other hand, large variations of these amounts are expected according to the particular history and properties of each object and the consequent changes in circumstellar conditions and chemistry. Hence the great variety of the relative intensities of the two FIR bands in observed PPN spectra. Changing the relative number of the structures selected in this work, or new ones for that matter (thus adding more free parameters), allows one to tailor the overall spectrum at will.

While sulphur may be found in a huge variety of molecules, it seems to carry the 21-$\mu$m band only when in the form of thiourea. The latter has a quite specific geometry, and, therefore, cannot be expected to occur under any circumstances. Thus, turbulent mixing in the envelope is expected to enhance the population of NH$_{3}$ and sulphur-bearing species (see for instance Heinzeller et al. \cite{heinz}).

      By contrast, the 30-$\mu$m band is carried by structures which have much less specific geometries (aliphatic chains) and may therefore be present under less stringent conditions. Hence the rarity of the 21-$\mu$m band, and its absence in cases where the other one is observed. Clearly, a simpler molecular structure or a solid carrier (such as TiC)  would be less elusive. The prerequisites for the existence of thiourea have probably more to do with atmospheric conditions and complicated chemical reaction paths rather than with gross relative abundances. This is typical food for molecular chemists. \rm

\begin{table*}
\caption[]{Elemental composition of dust}
\begin{flushleft}
\begin{tabular}{lllllll}
\hline
Name & N$_{at}$ & C & H & O & N & S \\
\hline
thiourea & 8 & 1 & 4 & 0 & 2 & 1 \\
\hline
compact deriv. & 131 & 64 & 54 & 0 & 8 & 5 \\
\hline
linear deriv. & 217 & 115 & 84 & 0 & 12 & 6  \\
\hline
aliph. chains & 381 & 109 & 224 & 48 & 0 & 0 \\
\hline
aliph chains & 291 & 83 & 176 & 32 & 0 & 0 \\
\hline
total & 1028 & 372 & 542 & 80 & 22 & 12 \\
\hline
\% & 100 & 36 & 53 & 8 & 2 & 1 \\
\hline

\end{tabular}
\end{flushleft}
\end{table*}

\section{Conclusion}

We have shown that the thiourea functional group, associated with various carbonaceous structures, has one or two strong emission lines in a spectral range of $\sim$4 $\mu$m, within the 21-$\mu$m band emitted by a number of pre-planetary nebulae. The combination of nitrogen and sulphur in thiourea is the essential source of emission in this model: the band disappears if these species are replaced by carbon. These two elements are part of the ubiquitous CHONS family because of their high chemical activity. Thiourea may therefore readily form in space, and be found as an independent molecule or as a peripheral group attached to carbonaceous structures believed to be abundant in space. In all cases, it carries a strong IR line near the molecular thiourea line, which is the strongest and thus determines the peak of the band.

Obviously, no single structure can exhibit the required spectrum, for each only contributes discrete lines which cannot be broadened enough by usual broadening mechanisms. Twelve structures have been selected here, but their list is far from being exhaustive; they are only intended as examples of the generic thiourea class. The chemical software used here also allows to determine the types of modal vibrations which cary the lines of interest; this helps designing new structures to fill the wide bands observed in the sky.

Using interpolation and smoothing between the concatenated discrete lines of the selected structures, we produced synthetic spectra which exhibit a prominent, asymmetric, feature between 18 and 25 $\mu$m, with half-maximum points at 19.6 and 21.9 $\mu$m, very near the observed values. However, the peak is 0.6 $\mu$m redward of the observed average. 

The astronomical 21-$\mu$m feature extends redward to merge with the other, prominent 30-$\mu$m band. It is found that the main characters of this band can be modelled by the combined spectra of: a) aliphatic chains, made of CH$_{2}$ groups, oxygen bridges and OH groups, which provide the 30-$\mu$m emission; 
b) small, mostly linear, aromatic structures, which contribute to raise the red wing of the 21-$\mu$m band and fill the space between the two main features. The concatenated spectral lines of ten of these structures form a strong band between 23 and 38 $\mu$m.

The omission of oxygen in such structures all but extinguishes the 30-$\mu$m emission. The fact that these carriers do not involve, and are likely more abundant than, thiourea derivatives ensures that the 30-$\mu$m feature can still be present in the absence of the 21-$\mu$m feature, as observed.

Combining the discrete lines of the 22 selected structures in different proportions, interpolating and smoothing, we produced 2 synthetic spectra which purport to mimic, respectively, 2 typical PPN spectra. The main defect of these model spectra is insufficient intensity near 27 $\mu$m, resulting in a small redshift of the ``30-$\mu$m" band peak.

As expected, the selected structures also contribute lines in the near- and mid- IR bands (UIBs), due to the various vibrations of their CC and CH bonds. To my knowledge, none of the lines displayed in the figures above has been shown to be absent in PPNe. Some of them could help confirming the presence of thiourea or derivatives thereof: e.g. the NH stretching lines near 3 $\mu$m (see also Stewart \cite{ste}), \bf although this might prove to be difficult in emission because of the low temperatures indicated by the fitting procedure.\rm

Finally, the mechanism of the commonly observed disappearance of the 21-$\mu$m line is straightforward, if this line is due to thiourea: indeed the corresponding vibration modes mainly involve the S,C,N and H atoms in a very special conformation, and the latter are easily removed from the structure. Thus, thiourea, by contrast with solid carriers or strongly bonded diatomic molecules, may be muted by radiation or H atom encounters, for instance.

\section{Appendix}

\begin{figure}
\resizebox{\hsize}{!}{\includegraphics{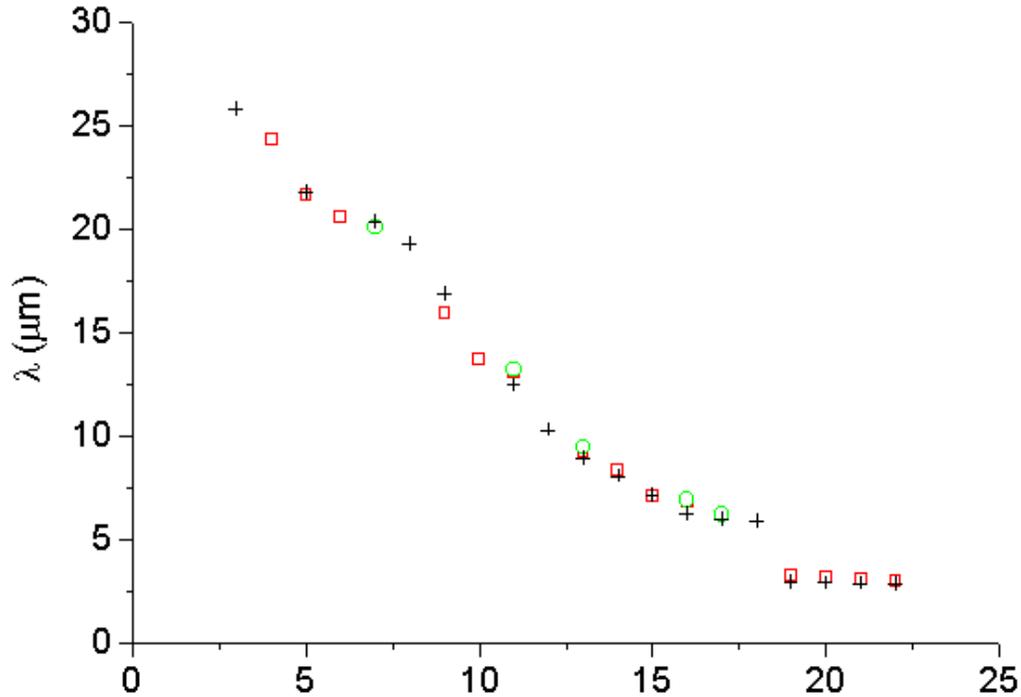}}
\caption[]{Vibrational wavelengths computed by Stewart \cite{ste} (red squares), Yamagushi et al. \cite{yam} (green circles) and in the present work (black crosses). \bf{The abscissa represents the rank of the lines by order of increasing wavenumber}\rm.}
\end{figure}

While the SCN$_{2}$ skeleton of thiourea is robustly framed into a plane, the two amine (NH$_{2}$) planes easily rotate about the CN axes or depart slightly from the skeletal plane. Hence the existence of conformeres studied theoretically by several authors (e.g. Bryantsev and Hay \cite{bry}).
A conformational search with the present chemical software also delivered four other conformeres: one of them is planar (C$_{2v}$ symmetry), one is slightly skewed out of plane like an S, as in Fig. 2, one is slightly skewed in the other sense (like a Z, also of symmetry C$_{2}$), and one of symmetry C$_{s}$ or TS1), with one pair of NH's in the plane NCN and the other oop. Other chemical algorithms yield similar results (see Bryantsev and Hay \cite{bry}, who provide clear representations of the geometrical differences between structures). The structural and energetic differences between the 4 structures are so small that they do not impede transitions from one to the other, and cause only minor changes in the vibrational modes. The corresponding frequencies for our conformeres are about 496, 492, 496 and 492 cm$^{-1}$, covering the range 20.1 to 20.4 $\mu$m. We therefore retained only one structure (Fig. 2).

Figure 14 displays the lines calculated by Stewart \cite{ste} and Yamagushi et al. \cite{yam}, together with those obtained in this work.

\section{Acknowledgments}
\bf It is a pleasure to acknowledge the useful suggestions of the reviewer, Dr Th. Posch, and the help of Prof. Sun Kwok in providing the data files for the PPNe.\rm

Note added in proof

After completion of this work, it was realized that the thiourea group in space could also be attached to aliphatic chains. The families of thiourea derivatives illustrated in Sec. 2 were therefore complemented with several structures comprizing one thiourea group attached to one of the carbon atoms along an aliphatic chain, including the ends. Again, the thiourea feature emerged in the vicinity of 20 $\mu$m (19.6 to 20.4 $\mu$m) with considerable intensity.

Further, investigation of various more complex and larger structures showed that the specific thiourea feature again appeared within the right wavelength range, but with decreasing intensity as the size of the structure increased.

\begin{thebibliography}{}


\bibitem[1999]{ali}Alia J. Edwards H. and Stoev M. 1999, Spectrochimica Acta A 55, 2423
\bibitem[1983]{boh}Bohren C. and Huffman D. 1983, Absorption and Scattering of Light by Small Particles, Wiley and Sons, New York
\bibitem[2006]{bre}Brennan N. 2006, Thesis, Univ. Pretoria, upeted.up. ac. za/thesis.       
\bibitem[2006]{bry}Bryantsev V. and Hay B. 2006, J. Phys. Chem. A, 110, 4678
\bibitem[2003]{chi}Chigai T., Yamamoto, Kaito C. and Kimura Y. 2003, ApJ 587, 771
\bibitem[1985]{goe85}Goebel J. and Moseley S. 1985,ApJ 290, L35
\bibitem[1993]{goe93}Goebel J. A\&A, 278, 226
\bibitem[1999]{hei}v. Heinjsbergen  1999, Phys. Rev. Lett. 83, 4986
\bibitem[2011]{heinz}Heinzeller D., Nomura H., Walsh C. and Millar T. 2011, arXiv:1102.3972v1
\bibitem[2001]{hen}Henning Th. and Mutschke H. 2001, Spectrochimica Acta A 57, 815
\bibitem[2002]{hon}Hony S., Waters L. and Tielens A. 2002, A\&A 390, 533
\bibitem[2009]{hri}Hrivnak B., Volk K. and Kwok S. 2009, Ap.J. 694, 1147
\bibitem[1961]{kut}Kutzelnigg W. and Mecke R. 1961, Spectrochimica Acta 17, 530
\bibitem[1989]{kwo89}Kwok S., Volk K. and Hrivnak B. 1989, ApJ 535, 275
\bibitem[2004]{les}Lesarri A., Mata S., Blanco S., Lopez J and Alonso J. 2004, J. Chem. Phys. 120, 6191
\bibitem[2003]{li}Li A. 2003, ApJ Lett. 
\bibitem[2009]{lin}Lin-Vien D., Colthup N., Fateley W. and Grasselli J. 1991, The Handbook of Infrared and Raman Characteristic Frequencies of Organic Molecules, Academic Press, New York.
\bibitem[2000]{mas}Masunov A. and Dannenberg J. 2000, J. Phys. Chem. B 104, 806
\bibitem[2009]{mer}Merck's FT-IR Atlas 1988, VCH, Germany
\bibitem[1995]{omo}Omont A., Moseley S., Cox P. et al. 1995, ApJ 454, 819
\bibitem[2000]{pap00}Papoular R. 2000, A\&A Lett. 362, L9
\bibitem[2001]{pap01}Papoular R. 2001, Spectrochimica Acta A 57, 947
\bibitem[2010]{pap}Papoular R. 2010, arXiv 1008.5136
\bibitem[2004]{pos}Posch Th., Mutschke H. and Andersen A. 2004, ApJ 616, 1167
\bibitem[1999]{smi} Smith B.  1999, Infrared Spectral Interpretation, CRC Press, New York.
\bibitem[2004]{spe}Speck A. and Hofmeister A. 2004, ApJ 600, 986
\bibitem[1957]{ste}Stewart J. 1957, J. Chem. Phys. 26, 248
\bibitem[1992]{sou}Sourrisseau C., Coddens G. and Papoular R. 1992, A\&A 254, L1
\bibitem[1999]{szc}Szczerba R., Henning Th., Volk K., Kwok S. and Cox P. 1999, A\&A 99, 345, L39
\bibitem[1987]{tur87a}Turner B. 1987, A\&A, 183, L23
\bibitem[1987]{tur87b}Turner B. 1987, A\&A, 183, L15
\bibitem[2010]{ull}Ullman's Encyclopedia of Industrial Chemistry 2010, Wiley-VCH 
\bibitem[2000]{hel}von Helden G., Tielens A., van Heijnsbergen D. et al. 2000, Science, 288, 313
\bibitem[1958]{yam}Yamagushi A. et al. 1958, J. Am. Chem. Soc. 80, 527
\bibitem[2010]{zhak}Zhang K., Jiang B. and Li A. 2009, MNRAS
\bibitem[2010]{zha}Zhang Y., Kwok S. and Hrivnak B. 2010, ApJ 725, 990
\end{thebibliography}
\end{document}